\documentclass{article} 
\usepackage{conference,times}

\usepackage{amsmath,amsfonts,bm}

\def\eqref#1{equation~\ref{#1}}

\def\1{\bm{1}}

\DeclareMathAlphabet{\mathsfit}{\encodingdefault}{\sfdefault}{m}{sl}
\SetMathAlphabet{\mathsfit}{bold}{\encodingdefault}{\sfdefault}{bx}{n}

\newcommand{\ourtool}{MCTS-SFT}
\usepackage[
    colorlinks=false,
    linkbordercolor=red,
    citebordercolor=green,
    urlbordercolor=cyan
]{hyperref}
\usepackage{url}
\usepackage{booktabs}
\usepackage{multirow}
\usepackage{adjustbox}

\title{Search-Guided Reasoning Trace Curation \\ for Verilog Generation}

\author{
\begin{tabular}[t]{@{}l@{}}
\textbf{Matthew DeLorenzo} \\
\normalfont Texas A\&M University \\
\normalfont \texttt{matthewdelorenzo@tamu.edu}
\end{tabular}
\And
\begin{tabular}[t]{@{}l@{}}
\textbf{Kevin Tieu} \\
\normalfont Texas A\&M University \\
\normalfont \texttt{kevin.tieu@tamu.edu}
\end{tabular}
\And
\begin{tabular}[t]{@{}l@{}}
\textbf{Prithwish Jana} \\
\normalfont Georgia Institute of Technology \\
\normalfont \texttt{pjana7@gatech.edu}
\end{tabular}
\And
\begin{tabular}[t]{@{}l@{}}
\textbf{Piyush Jha} \\
\normalfont Georgia Institute of Technology \\
\normalfont \texttt{piyush.jha@gatech.edu}
\end{tabular}
\And
\begin{tabular}[t]{@{}l@{}}
\textbf{Dileep Kalathil} \\
\normalfont Texas A\&M University \\
\normalfont \texttt{dileep.kalathil@tamu.edu}
\end{tabular}
\And
\begin{tabular}[t]{@{}l@{}}
\textbf{Vijay Ganesh} \\
\normalfont Georgia Institute of Technology \\
\normalfont \texttt{vganesh45@gatech.edu}
\end{tabular}
\And
\begin{tabular}[t]{@{}l@{}}
\textbf{Jeyavijayan Rajendran} \\
\normalfont Texas A\&M University \\
\normalfont \texttt{jv.rajendran@tamu.edu}
\end{tabular}
}

\iclrfinalcopy 
\begin{document}

\maketitle

\begin{abstract}
Recent advances in reasoning-capable large language models (LLMs) have improved Verilog generation. However, most training pipelines distill natural-language reasoning traces whose intermediate design decisions are weakly structured and are primarily selected according to the correctness of the final generated RTL. We present \textbf{MCTS-SFT}, a verifier-guided framework for curating reasoning traces for supervised fine-tuning (SFT) of Verilog generation models. MCTS-SFT formulates trace construction as search over intermediate reasoning states and uses Monte Carlo Tree Search (MCTS) to propagate functional feedback from complete Verilog rollouts to earlier design decisions. Within this framework, we introduce \textbf{Plan-MCTS}, which searches over compact hardware-design plans that explicitly represent interfaces, timing, state, behavior, and update rules. We further study \textbf{plan-preserving supervision}, in which the searched structured state is retained as part of the downstream SFT target. Our experiments find that Plan-MCTS achieves the highest functional trace yield, with 84.2\% correct traces compared with 81.9\% for one-shot structured planning and 82.0\% for unguided tree exploration. For identical Plan-MCTS trajectories and Verilog outputs, retaining the explicit searched plan during SFT raises pass@10 from 52.56\% to 57.7\%, demonstrating that the representation of intermediate design decisions materially affects downstream learning. In a separate 6,000-trace study, MCTS-curated supervision improves pass@10 over full-response chain-of-thought distillation by up to 3.7 points while reducing downstream generation length by approximately 50\% on VerilogEval-v2. These results demonstrate that structured hardware planning, verifier-guided search, and explicit intermediate-state supervision provide complementary mechanisms for constructing more effective and token-efficient training data for Verilog generation.
\end{abstract}

\section{Introduction}

Large language models (LLMs) have recently made substantial progress on long-horizon, multi-step tasks, driven in part by reasoning models that generate extended intermediate traces before producing a final response~\citep{guo2025deepseek,wang2024math}. These reasoning trajectories have become an increasingly important source of supervision for distillation and fine-tuning, enabling smaller models to inherit aspects of the reasoning behavior of stronger teachers. However, the quality of the intermediate traces remains underexplored. Long reasoning sequences can introduce substantial token overhead and may contain repetitive, inconsistent, or weakly relevant content~\citep{qu2025survey}, motivating the construction of effective reasoning supervision.

This issue is particularly relevant for hardware description language (HDL) generation. Recent work has applied reasoning-capable LLMs to functional Verilog generation and has shown that fine-tuning on chain-of-thought (CoT) responses, often combined with executable feedback, can substantially improve final design correctness~\citep{yubeaton2025verithoughts,zhu2025qimeng,wang2025verireason}. Existing approaches, however, generally curate complete reasoning responses and evaluate them according to the final generated Verilog, providing limited control over the quality of the intermediate reasoning that is ultimately used for training. Moreover, these traces are typically expressed as free-form natural language, despite RTL design being naturally organized around explicit decisions concerning interfaces, timing, state, behavior, and signal updates. This motivates a reasoning-trace curation process that can both evaluate intermediate design decisions using functional feedback and represent them in a form that provides effective supervision for downstream fine-tuning.

To address this problem, we present \textbf{MCTS-SFT}, a verifier-guided framework for curating Verilog reasoning traces for supervised fine-tuning. MCTS-SFT formulates intermediate reasoning as a sequence of design states and applies Monte Carlo Tree Search (MCTS) to estimate their quality from terminal testbench feedback. We study two forms of intermediate representation. \textit{Thought-MCTS} searches directly over free-form reasoning steps, while \textit{Plan-MCTS} maintains a compact structured hardware plan whose updates correspond to explicit design decisions. The resulting search trajectories are then converted into prompt-response examples for SFT, allowing the student model to learn from reasoning paths selected using executable feedback. In Plan-MCTS, we additionally retain the searched plan itself as part of the training response, directly exposing the intermediate hardware decisions discovered during curation.

Our main contributions are:
\begin{enumerate}
    \item We introduce \textbf{MCTS-SFT}, a verifier-guided framework for curating reasoning traces for Verilog supervised fine-tuning. MCTS-SFT uses terminal testbench feedback to assign value to intermediate reasoning states, enabling executable verification to guide training-data construction without a separate process-reward model.

    \item Within MCTS-SFT, we introduce \textbf{Plan-MCTS}, a structured curation strategy that represents intermediate reasoning as explicit hardware design decisions rather than accumulated free-form text. Controlled comparisons show that structured planning is independently useful, while verifier-guided search provides additional benefit for discovering functionally correct training trajectories.

    \item We demonstrate that \textbf{preserving the searched hardware plan as SFT supervision} is critical to downstream learning. For identical Plan-MCTS trajectories and Verilog outputs, explicit plan supervision improves pass@10 from 52.56\% to 57.7\%; in the larger 6k-trace study, MCTS-curated SFT also improves correctness over full-response CoT distillation while reducing generation length by approximately 50\% on VerilogEval-v2.
\end{enumerate}

\section{Background}

\subsection{Training LLMs for Verilog Generation}

Specializing LLMs for Verilog generation has become an active research area as a means of addressing limitations in hardware design capabilities~\citep{yang2025large}. Early work focused on domain-specific pretraining and supervised fine-tuning over curated Verilog corpora~\citep{thakur2024verigen}, followed by instruction-tuning approaches that expanded prompt-response datasets through filtering and augmentation strategies~\citep{zhao2024codev, liu2024rtlcoder}. More recent methods incorporate functional feedback directly into post-training. VeriRL applies reinforcement learning using similarity to reference implementations~\citep{teng2025verirl}, while subsequent work introduced module-specific testbench generation to enable functionality-based rewards~\citep{wang2025insights}. Other approaches have explored more specialized reward formulations, including hierarchical~\citep{chen2025chipseek} and entropy-aware objectives~\citep{shi2025earl}.

\subsection{Reasoning LLMs for Verilog Generation}

The emergence of reasoning LLMs~\citep{guo2025deepseek} has motivated training pipelines that incorporate explicit reasoning traces into Verilog generation. VeriThoughts~\citep{yubeaton2025verithoughts} distills DeepSeek-R1 reasoning traces for SFT after verifying the resulting Verilog against reference solutions. QiMeng-CodeV-R1~\citep{zhu2025qimeng} similarly constructs reasoning-based training data with functional checking and applies DAPO reinforcement learning, while VeriReason~\citep{wang2025verireason} uses GRPO with dataset refinement and structural Verilog rewards. These approaches substantially improve functional correctness, with CodeV-R1 reporting 12--20\% gains over prior state-of-the-art results on VerilogEval~\citep{zhu2025qimeng}, but also expose challenges including long and inefficient reasoning trajectories~\citep{wang2025verireason, yubeaton2025verithoughts}. ReasoningV adapts the amount of reasoning to problem complexity~\citep{qin2025reasoningv}, while SiliconMind-V1 introduces a multi-agent refinement process that incorporates LLM-generated testbench feedback into SFT without external EDA tools~\citep{chen2026siliconmind}. QiMeng-CRUX further introduces a structured thought representation for GRPO, using confidence-based rewards to optimize intermediate reasoning structure~\citep{huang2026qimeng}. Independent analysis of RTL reasoning traces also identifies substantial repetition and token overhead~\citep{delorenzo2025tracing}. 
Recent work has also begun to explicitly assign value to intermediate RTL reasoning steps. StepPRM-RTL~\citep{vijayaraghavan2026stepprm} represents RTL construction as a sequence of rationale--code-edit pairs and trains a step-level process reward model (PRM) to score intermediate decisions. The learned PRM is then used to guide MCTS exploration of alternative reasoning trajectories, with the resulting trajectories incorporated into a retrieval-augmented fine-tuning procedure. This provides process-level credit assignment during RTL reasoning, but requires learning an auxiliary reward model from intermediate steps.

Our work addresses this problem from a complementary direction. Rather than training a separate process reward model, \ourtool{} uses executable feedback from complete Verilog rollouts as the sole source of search value and propagates this terminal feedback through MCTS to earlier reasoning states. Moreover, Plan-MCTS represents these states as persistent, hardware-specific design plans and studies the searched state representations that should be retained as downstream SFT supervision. Therefore, our focus is on using terminal functional verification to discover and value intermediate design states without requiring direct process-level labels or a learned verifier.
\subsection{Inference-Level Verilog Reasoning Optimization}

A complementary line of work improves Verilog generation at inference time through structured prompting, candidate selection, or iterative refinement. Abstractions-of-Thought~\citep{delorenzo2025abstractions} uses a staged prompting procedure that transitions from higher-level to lower-level intermediate representations. VRank~\citep{zhao2025vrank} clusters sampled Verilog candidates according to simulated behavior and selects outputs using self-consistency, while VFocus~\citep{zhao2025vfocus} combines CoT sampling, reasoning-length filtering, simulation-based ranking, and refinement of promising candidates. Other agentic and feedback-driven systems similarly use compiler diagnostics, simulation feedback, generated testbenches, or iterative repair, while evolutionary approaches refine candidate implementations with respect to correctness or hardware metrics~\citep{min2026revolution, ping2026coevo, hsin2026evolve}. These methods demonstrate the value of search and reasoning structure for improving final RTL candidates, but primarily operate at inference time. In contrast, \ourtool{} uses verifier-guided search offline to explore intermediate reasoning sequences, curate training trajectories, and leverage them for supervised fine-tuning.

\begin{figure}[t]
    \centering
    \includegraphics[width=1.0\textwidth]{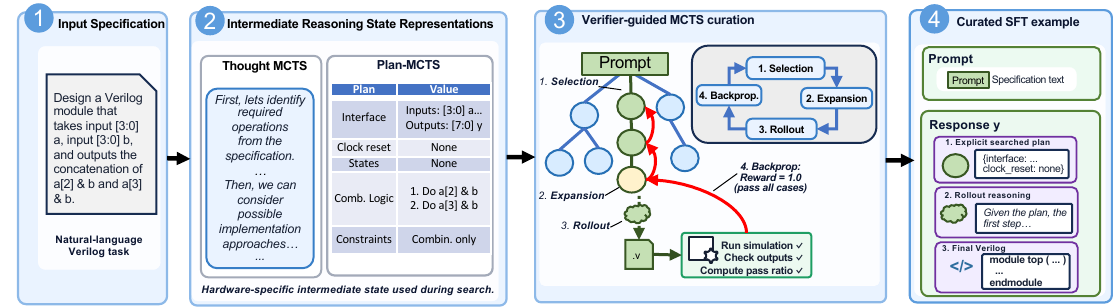}
    \caption{MCTS-SFT trace curation process, including state representations and MCTS stages.}
    \label{fig:my_image}
\end{figure}

\section{Framework}
\label{sec:framework}

Given a natural-language Verilog specification $x$, \ourtool{} constructs reasoning supervision by using executable feedback to evaluate and curate intermediate design decisions before supervised fine-tuning. For each specification, a teacher model explores intermediate reasoning states, selected states are rolled out into complete reasoning traces and Verilog implementations, and the resulting code is evaluated by a functional verifier. Monte Carlo Tree Search (MCTS) propagates terminal verification feedback to earlier reasoning states, allowing otherwise non-executable design decisions to influence subsequent exploration. The resulting trajectories are then converted into prompt-response examples for fine-tuning a student model.

\ourtool{} denotes the complete curation and fine-tuning framework. Within this framework, we consider two intermediate-state representations. \textit{Thought-MCTS} searches over accumulated free-form reasoning, while \textit{Plan-MCTS} searches over structured hardware design states. Our primary formulation uses Plan-MCTS and retains the searched hardware plan as part of the downstream SFT supervision. Figure~\ref{fig:my_image} summarizes the complete pipeline.

\subsection{Verifier-Guided Reasoning-Trace Curation}
\label{sec:verifier_search}

We formulate reasoning-trace construction for a specification $x$ as a finite-horizon decision process
\begin{equation}
    \mathcal{M}_x = (\mathcal{S},\mathcal{A},T,R),
\end{equation}
where a state contains the specification together with its current intermediate reasoning representation, an action modifies that representation, $T$ defines the corresponding transition, and $R$ is obtained by verifying a complete Verilog rollout. At reasoning step $t$,
\begin{equation}
    S_t=(x,h_t),
    \qquad
    S_{t+1}=T(S_t,a_t),
    \label{eq:state_transition}
\end{equation}
where $h_t$ denotes the reasoning state constructed after $t$ decisions. The interpretation of $h_t$ depends on the search representation: Thought-MCTS uses an accumulated natural-language reasoning prefix, whereas Plan-MCTS uses the structured hardware plan introduced in Section~\ref{sec:plan_mcts}.

A central difficulty is that an intermediate state cannot be functionally verified in isolation. Decisions concerning timing, state organization, or control logic acquire executable meaning only after they are instantiated in a complete RTL implementation. We therefore evaluate an intermediate state through a rollout model,
\begin{equation}
    (r,v)
    \sim
    \pi_{\theta}^{\mathrm{rollout}}
    (\cdot \mid x,h_t),
    \label{eq:rollout}
\end{equation}
which materializes the current state into a complete reasoning trace $r$ and Verilog implementation $v$. The implementation is then compiled and evaluated against the corresponding testbench, producing terminal reward
\begin{equation}
    R(v)=
    \begin{cases}
        \alpha_{\mathrm{NC}}, & \text{if $v$ does not compile}, \\[3pt]
        \dfrac{N_{\mathrm{pass}}}{N_{\mathrm{total}}}, & \text{otherwise},
    \end{cases}
    \label{eq:reward}
\end{equation}
where $\alpha_{\mathrm{NC}}<0$ penalizes non-compilable designs and $N_{\mathrm{pass}}/N_{\mathrm{total}}$ measures functional correctness.

\paragraph{Verifier-guided tree search.}
MCTS repeatedly performs selection, expansion, rollout, and backpropagation over this intermediate-state space. During selection, the next explored action is chosen according to the UCT criterion
\begin{equation}
    a^{*}
    =
    \arg\max_{a \in \mathcal{A}_{\mathrm{tree}}(S)}
    \left[
        Q(S,a)
        +
        c_{\mathrm{uct}}
        \sqrt{
            \frac{\log N(S)}
                 {1+N(S,a)}
        }
    \right],
    \label{eq:uct}
\end{equation}
where $Q(S,a)$ estimates the downstream functional value of action $a$, $N(S)$ and $N(S,a)$ denote state and edge visitation counts, and $c_{\mathrm{uct}}$ controls exploration.

When an expandable state is reached, the teacher proposes multiple candidate next decisions, each defining a successor state through Eq.~\ref{eq:state_transition}. A selected successor is rolled out according to Eq.~\ref{eq:rollout}, and its terminal verifier reward is propagated through the visited path. We maintain a running action-value estimate
\begin{equation}
    Q(S,a)
    \leftarrow
    Q(S,a)
    +
    \frac{R(v)-Q(S,a)}
         {N(S,a)}.
    \label{eq:q_update}
\end{equation}
Repeated simulations use executable feedback to score complete implementations and guide subsequent exploration toward intermediate decisions associated with stronger functional outcomes. Additional search details, pseudocode, and generation parameters are provided in Appendix~\ref{app:mcts_details}.

\subsection{Plan-MCTS: Structured Hardware Design States}
\label{sec:plan_mcts}

Plan-MCTS instantiates the intermediate state $h_t$ as a structured hardware design plan rather than an accumulated prose trace. The plan contains seven hardware-specific fields:
\texttt{interface},
\texttt{clock\_reset},
\texttt{state\_elements},
\texttt{combinational\_logic},
\texttt{fsm},
\texttt{edge\_cases}, and
\texttt{implementation\_constraints}.
Together, these fields capture the principal architectural and behavioral decisions required to translate the natural-language specification into RTL while remaining independent of the exact syntax of the final implementation.

At step $t$, the current plan is denoted by $p_t$. An expansion action proposes a structured update
\begin{equation}
    p_{t+1}
    =
    \mathrm{Update}(p_t,\Delta p_t),
    \label{eq:plan_update}
\end{equation}
where $\Delta p_t$ identifies a target field, proposes a design decision, and provides a concise rationale. The teacher determines which aspect of the current plan should be refined and generates alternative updates during expansion. Consequently, branches in the search tree correspond to alternative hardware design decisions rather than alternative continuations of unrestricted natural-language reasoning.

The structured plan is intentionally an intermediate design representation rather than executable RTL. A selected plan is therefore evaluated by conditioning the rollout model on both the specification and accumulated plan,
\begin{equation}
    (r,v)
    \sim
    \pi_{\theta}^{\mathrm{rollout}}
    (\cdot \mid x,p_t),
    \label{eq:plan_rollout}
\end{equation}
after which the resulting Verilog implementation is functionally verified and its reward is backpropagated through the sequence of plan updates.

This gives the plan two complementary roles within \ourtool{}. First, it defines a hardware-specific state space in which MCTS can explore and assign value to intermediate design choices. Second, it provides an explicit record of the decisions associated with the resulting implementation, which can subsequently be exposed to the student during fine-tuning.

For comparison, Thought-MCTS applies the same verifier-guided search procedure while representing $h_t$ as an accumulated free-form reasoning prefix. This isolates the effect of structured hardware representation from the effect of MCTS itself. Likewise, Plan-CoT and Plan-RandomTree introduced in Section~\ref{sec:experiments} distinguish structured planning from verifier-guided search and from branching without reward-guided selection, respectively.

\subsection{Plan-Preserving Supervised Fine-Tuning}
\label{sec:plan_sft}

For a curated Plan-MCTS trajectory, let $(x,p,r,v)$ denote the input specification, searched hardware plan, rollout reasoning, and terminal Verilog implementation. The corresponding training target is
\begin{equation}
    y^{\mathrm{plan}}
    =
    \mathrm{Serialize}(p,r,v),
    \label{eq:sft_target}
\end{equation}
where $\mathrm{Serialize}(\cdot)$ retains the explicit field structure of $p$ together with the materialized reasoning and final implementation. The resulting $(x,y^{\mathrm{plan}})$ pairs are used for standard autoregressive SFT of the student model.

This final stage separates two distinct aspects of reasoning-trace curation. \emph{Search quality} determines which intermediate design states and terminal implementations are discovered through verifier-guided exploration, while \emph{supervision quality} determines how those states are represented to the downstream model. Our controlled experiments separately evaluate these effects by comparing planning with and without verifier-guided search and, for fixed Plan-MCTS trajectories, varying only the intermediate representation retained in the SFT response. Additional details relating to the SFT training configurations are included in the Appendix.

\section{Experiments}
\label{sec:experiments}

We evaluate \ourtool{} through controlled and extended-scale experiments addressing three questions:
\textbf{RQ1:} How do structured planning and verifier-guided search affect the discovery of functional training trajectories?
\textbf{RQ2:} How does the intermediate representation retained during SFT affect downstream learning?
\textbf{RQ3:} How does search-curated reasoning supervision affect downstream correctness, generation efficiency, and transfer at larger scale?

We use two complementary protocols. A \textbf{controlled 1k-trace study} isolates the effects of structured planning, verifier-guided search, and SFT-target representation under a unified teacher model. An \textbf{extended 6k-trace study} evaluates the broader MCTS-SFT pipeline across multiple Verilog benchmarks.

\subsection{Experimental Setup}

\paragraph{Data and verification.}
We construct training examples from natural-language Verilog specifications sampled from the VeriThoughts training set~\citep{yubeaton2025verithoughts}. Generated implementations are compiled and evaluated using automatically constructed testbenches following prior Verilog reasoning work~\citep{zhu2025qimeng}. A trajectory is considered functionally correct when its terminal Verilog implementation passes all associated testbench cases.

\paragraph{Controlled 1k-trace study.}
We sample 100 specifications and generate 10 trajectories per specification for each curation strategy, yielding 1,000 examples per dataset. DeepSeek-R1 is used for every LLM invocation, and all methods share the same verifier, generation budget, student model, and SFT configuration. This matched setting is used for the component-level comparisons in RQ1 and RQ2.

We compare five curation strategies. \textbf{CoT} independently samples complete free-form reasoning responses. \textbf{Thought-MCTS} applies verifier-guided search over free-form reasoning states. \textbf{Plan-CoT} generates a complete structured hardware plan in a single step before rollout, providing a planning baseline without tree search. \textbf{Plan-RandomTree} matches the structured state representation, branching process, rollout procedure, and search budget of Plan-MCTS, but selects nodes uniformly rather than using verifier-guided values. \textbf{Plan-MCTS} incrementally constructs the structured hardware plan using the search procedure in Section~\ref{sec:framework}. Together, these controls separate the effects of structured representation, tree exploration, and verifier-guided selection.

For RQ2, we additionally construct three SFT targets from identical Plan-MCTS trajectories: \textbf{rollout only}, which omits the searched plan; \textbf{rendered plan}, which prepends a natural-language rendering of the plan; and \textbf{explicit plan}, which preserves the structured hardware representation directly. Since the underlying trajectories, terminal Verilog implementations, rewards, and dataset sizes are fixed, this comparison isolates the effect of SFT-target representation. We also include \textbf{Summary-MCTS}, which compresses accumulated free-form reasoning before subsequent expansion, as a control for reasoning verbosity.

\paragraph{Extended 6k-trace study.}
We additionally evaluate a larger-scale setting using 600 specifications and 10 trajectories per specification, yielding 6,000 examples for each full curation strategy. We compare CoT, Thought-MCTS, and Plan-MCTS, together with answer-only SFT and a 600-example Best-of-$N$ Plan-MCTS baseline.

This experiment utilizes a mixed teacher configuration: DeepSeek-R1 is used for CoT generation, Thought-MCTS, and Plan-MCTS rollouts, while DeepSeek-V3 constructs intermediate Plan-MCTS states. It also uses a \emph{rollout-only} Plan-MCTS SFT target rather than the explicit plan-preserving target studied in the controlled experiment. We therefore use the 6k study for pipeline-level comparisons of downstream accuracy and generation efficiency, while component-level conclusions are drawn from the unified-teacher 1k study.

\paragraph{Supervised fine-tuning.}
For all datasets, we fine-tune the same 7B Qwen2.5 student model using an identical SFT procedure. We train for three epochs with learning rate $5\times10^{-5}$, maximum sequence length 16,384, and LoRA rank 64 on a single NVIDIA H200 GPU. Additional generation and training details are provided in Appendix~\ref{app:implementation}.

\paragraph{Evaluation.}
For the controlled study, we evaluate on VerilogEval-v1-Human~\citep{liu2023verilogeval} using $n=10$ generations per problem. The extended study additionally evaluates VerilogEval-v2 Code Completion and Spec-to-RTL using $n=20$ generations per problem. We report pass@$k$ functional correctness and average generated response length. For RQ1, we report \emph{functional trace yield}, defined as the proportion of curated trajectories whose terminal implementation passes all testbench cases.

To examine transfer beyond the predominantly single-module VerilogEval setting, we additionally evaluate on 172 non-agentic multi-module tasks from CVDP-002 and CVDP-003.

\section{Results}
\label{sec:results}

We organize the results around the three research questions in Section~\ref{sec:experiments}: the effects of planning and verifier-guided search during curation, the representation retained during SFT, and downstream performance at larger scale.

\subsection{RQ1: How Do Structured Planning and Verifier-Guided Search Affect Functional Trace Discovery?}
\label{sec:results_rq1}

Table~\ref{tab:rq1_curation} reports functional trace yield in the controlled 1k-trace setting, where all methods use the same teacher, specifications, generation budget, and verifier.

\begin{table}[t]
\centering
\small
\caption{
\textbf{RQ1: Functional trace discovery under the controlled 1k-trace setting.}
Each strategy generates 1,000 trajectories from the same 100 specifications using DeepSeek-R1 for all LLM calls. A correct trace passes all associated testbench cases.
}
\label{tab:rq1_curation}
\begin{tabular}{l l l r r}
\toprule
\textbf{Strategy}
& \textbf{State representation}
& \textbf{Exploration}
& \textbf{Correct}
& \textbf{Yield} \\
\midrule
CoT
& Free-form
& Independent
& 731 / 1,000
& 73.1\% \\

Thought-MCTS
& Free-form
& Verifier-guided
& 836 / 1,000
& 83.6\% \\

Plan-CoT
& Structured plan
& One-shot
& 819 / 1,000
& 81.9\% \\

Plan-RandomTree
& Structured plan
& Random tree
& 820 / 1,000
& 82.0\% \\

Plan-MCTS
& Structured plan
& Verifier-guided
& \textbf{842 / 1,000}
& \textbf{84.2\%} \\
\bottomrule
\end{tabular}
\end{table}

Through these results, we can see that structured planning alone results in improves functional trace yield. Plan-CoT reaches 81.9\% functional yield, compared with 73.1\% for independent CoT sampling. Random tree exploration provides essentially no additional benefit (82.0\%), whereas verifier-guided Plan-MCTS achieves the highest yield at 84.2\%. Thus, the hardware-specific representation provides a strong basis for trace construction, with verifier-guided selection contributing additional improvements over planning or branching alone.

Thought-MCTS also attains a high functional yield of 83.6\%, indicating that verifier-guided search can identify strong terminal implementations even with free-form reasoning states. However, curation-time correctness alone does not determine the usefulness of a trajectory for downstream learning, motivating the representation analysis in RQ2.

\subsection{RQ2: How Does Intermediate Representation Affect Downstream SFT?}
\label{sec:results_rq2}

Table~\ref{tab:rq2_representation} evaluates the controlled SFT datasets. The three Plan-MCTS target variants use identical search trajectories, Verilog implementations, rewards, and dataset sizes, isolating the effect of how the searched plan is presented to the student.

\begin{table}[t]
\centering
\small
\caption{
\textbf{RQ2: Effect of reasoning and SFT-target representation.}
All models are trained on 1,000 examples and evaluated on VerilogEval-v1-Human with $n=10$. The three Plan-MCTS target variants use identical search trajectories, terminal Verilog implementations, and verifier rewards.
}
\label{tab:rq2_representation}
\begin{tabular}{l r r r}
\toprule
\textbf{Method / SFT target}
& \textbf{pass@1}
& \textbf{pass@5}
& \textbf{pass@10} \\
\midrule
CoT
& 31.22
& 46.91
& 52.56 \\

Thought-MCTS
& 27.37
& 46.71
& 51.28 \\

Summary-MCTS
& 29.23
& 46.43
& 51.92 \\

Plan-CoT
& \textbf{32.37}
& 50.40
& 57.05 \\

Plan-RandomTree
& 28.53
& 47.44
& 53.21 \\
\midrule
Plan-MCTS, rollout only
& 29.17
& 46.48
& 52.56 \\

Plan-MCTS + rendered NL plan
& 30.00
& 47.60
& 55.10 \\

Plan-MCTS + explicit plan
& 31.50
& \textbf{50.80}
& \textbf{57.70} \\
\bottomrule
\end{tabular}
\end{table}

The fixed-trajectory comparison shows that supervision quality depends strongly on the retained intermediate representation. Plan-MCTS with rollout-only supervision obtains 52.56\% pass@10; adding a natural-language rendering of the plan raises this to 55.10\%, while retaining the explicit structured plan reaches 57.70\%. As the underlying trajectories and terminal programs are identical, the 5.14-point improvement from rollout-only to explicit-plan supervision is attributable to the intermediate information exposed during SFT. The explicit representation also outperforms the rendered plan at pass@1/5/10, indicating that preserving the structured hardware fields is more effective than expressing the same information as additional prose.

The remaining controls reinforce this conclusion. Plan-CoT reaches 57.05\% pass@10 without tree search, highlighting that structured planning is independently valuable, while explicit-plan Plan-MCTS obtains the strongest pass@5 and pass@10. In contrast, Summary-MCTS remains close to Thought-MCTS, suggesting that the benefit is not explained simply by compressing free-form reasoning.

\paragraph{Trace characteristics.}
Table~\ref{tab:trace_characteristics} further characterizes the controlled datasets.
\begin{table}[t]
\centering
\small
\caption{
Trace characteristics in the controlled study. Lower Distinct-2 indicates greater lexical repetition. The three Plan-MCTS target variants share the same terminal trajectories and therefore have the same mean verifier reward.
}
\label{tab:trace_characteristics}
\begin{tabular}{l r r r}
\toprule
\textbf{Strategy / target}
& \textbf{Mean reward}
& \textbf{Mean tokens}
& \textbf{Distinct-2} \\
\midrule
CoT
& 0.571
& 1,021
& 0.789 \\

Thought-MCTS
& 0.754
& 3,121
& 0.607 \\

Plan-CoT
& 0.755
& 1,649
& 0.747 \\

Plan-RandomTree
& 0.739
& 1,556
& 0.740 \\

Plan-MCTS, rollout only
& \textbf{0.780}
& 1,561
& 0.743 \\

Plan-MCTS + rendered NL plan
& \textbf{0.780}
& 1,681
& 0.745 \\

Plan-MCTS + explicit plan
& \textbf{0.780}
& 1,759
& 0.748 \\
\bottomrule
\end{tabular}
\end{table}
Thought-MCTS produces substantially longer and more repetitive traces than the plan-based methods despite its high curation yield. More importantly, the three Plan-MCTS targets share identical terminal Verilog implementations and verifier rewards, but markedly different downstream performance. This further reinforces the role of curated intermediate reasoning structures within the SFT procedure for effective Verilog generation.

\subsection{RQ3: Does Search-Curated Supervision Improve Downstream Correctness and Efficiency at Larger Scale?}
\label{sec:results_rq3}

We next evaluate the extended 6k-trace setting across VerilogEval-v1-Human and both VerilogEval-v2 tasks. As described in Section~\ref{sec:experiments}, this experiment uses a mixed-teacher configuration and rollout-only Plan-MCTS target, so we interpret it as a pipeline-level evaluation rather than a component attribution study.

\begin{table*}[t]
\centering
\small
\caption{
\textbf{RQ3: Downstream Verilog generation after SFT in the extended study.}
VE-v1-H denotes VerilogEval-v1-Human, VE-v2-CC denotes VerilogEval-v2 Code Completion, and VE-v2-S2R denotes VerilogEval-v2 Spec-to-RTL. Tokens report mean generated response length $\pm$ standard deviation. Bold denotes the highest pass@$k$ value for each benchmark.
}
\label{tab:rq3_verilogeval}
\resizebox{\textwidth}{!}{
\begin{tabular}{l r | r r r r | r r r r | r r r r}
\toprule
&
&
\multicolumn{4}{c|}{\textbf{VE-v1-H}}
&
\multicolumn{4}{c|}{\textbf{VE-v2-CC}}
&
\multicolumn{4}{c}{\textbf{VE-v2-S2R}} \\
\textbf{Training data}
& \textbf{\# SFT}
& \textbf{p@1}
& \textbf{p@5}
& \textbf{p@10}
& \textbf{Tokens}
& \textbf{p@1}
& \textbf{p@5}
& \textbf{p@10}
& \textbf{Tokens}
& \textbf{p@1}
& \textbf{p@5}
& \textbf{p@10}
& \textbf{Tokens} \\
\midrule
Baseline (no SFT)
& 0
& 21.4 & 40.9 & 46.6 & $1132 \pm 1640$
& 25.1 & 40.1 & 46.0 & $456 \pm 665$
& 24.1 & 38.6 & 43.7 & $422 \pm 649$ \\

CoT (answer only)
& 6,000
& 36.1 & 48.1 & 52.6 & $1888 \pm 4217$
& 35.7 & 50.1 & 54.4 & $442 \pm 466$
& 34.0 & 49.0 & 55.2 & $451 \pm 627$ \\

Best-of-$N$ (Plan-MCTS)
& 600
& 33.5 & 49.4 & 54.4 & $3143 \pm 3045$
& 35.3 & 50.7 & 55.4 & $3556 \pm 3178$
& 30.3 & 48.0 & 54.0 & $3760 \pm 3217$ \\

CoT (full response)
& 6,000
& 36.9 & 52.5 & 57.2 & $3162 \pm 3159$
& 41.5 & 54.8 & 58.8 & $7317 \pm 1757$
& 41.2 & 55.6 & 60.3 & $7313 \pm 1715$ \\

Thought-MCTS
& 6,000
& 35.9 & 53.5 & 57.8 & $4521 \pm 3057$
& 36.9 & 55.3 & 60.4 & $4348 \pm 3038$
& 41.3 & 55.8 & 60.3 & $4483 \pm 3017$ \\

Plan-MCTS
& 6,000
& \textbf{39.5}
& \textbf{55.7}
& \textbf{60.4}
& $2993 \pm 3173$
& \textbf{43.3}
& \textbf{57.4}
& \textbf{62.5}
& $3622 \pm 2962$
& \textbf{41.4}
& \textbf{55.9}
& \textbf{60.6}
& $3710 \pm 2973$ \\
\bottomrule
\end{tabular}
}
\end{table*}

We find that Plan-MCTS achieves the strongest functional performance across all three benchmarks. Relative to full-response CoT distillation, it improves pass@1/5/10 by 2.6/3.2/3.2 points on VerilogEval-v1-Human and by 1.8/2.6/3.7 points on VerilogEval-v2 Code Completion. Table \ref{tab:cvdp} indicates that gains on CVDP (Spec-to-RTL) problems are smaller, but remain positive at each reported pass@$k$.

A primary advantage of MCTS-SFT appears in \textit{generation efficiency}. As shown in Table~\ref{tab:rq3_verilogeval}, on the two VerilogEval-v2 prompt sets, the downstream response of models trained via CoT produces averages of 7,317 and 7,313 tokens, whereas Plan-MCTS reduces these to 3,622 and 3,710, corresponding to 50.5\% and 49.2\% reductions while maintaining or improving functional accuracy. Thought-MCTS is less consistent in accuracy and generates longer responses than Plan-MCTS across all three benchmarks. The answer-only and Best-of-$N$ controls likewise do not match the full Plan-MCTS dataset, although the smaller Best-of-$N$ training set prevents attributing this difference solely to its selection procedure.

\paragraph{Transfer to multi-module RTL generation.}
Finally, we evaluate transfer beyond the predominantly single-module VerilogEval setting using 172 non-agentic CVDP-002/003 tasks.
\begin{table}[t]
\centering
\small
\caption{
Preliminary multi-module evaluation on 172 CVDP-002/003 tasks. Absolute success rates remain low, and we therefore interpret this experiment as evidence of transfer rather than a comprehensive evaluation of multi-module RTL generation.
}
\label{tab:cvdp}
\begin{tabular}{l r r r}
\toprule
\textbf{Model}
& \textbf{Overall}
& \textbf{Easy}
& \textbf{Medium} \\
\midrule
CoT-SFT
& 9.30\% (16/172)
& 15.73\%
& 2.41\% \\

Thought-MCTS-SFT
& 11.63\% (20/172)
& \textbf{20.22\%}
& 2.41\% \\

Plan-MCTS-SFT
& \textbf{12.21\% (21/172)}
& 19.10\%
& \textbf{4.82\%} \\
\bottomrule
\end{tabular}
\end{table}
Plan-MCTS obtains the highest overall success rate at 12.21\%, compared with 11.63\% for Thought-MCTS and 9.30\% for CoT-SFT. On medium-difficulty tasks, it solves four examples compared with two for each baseline, although Thought-MCTS performs slightly better on the easy subset. Given the low absolute success rates of the student model, we treat these results as preliminary evidence of transfer rather than evidence of strong multi-module capability.

\section{Conclusion}

We present MCTS-SFT, a verifier-guided framework for curating reasoning traces for supervised fine-tuning of Verilog generation models. The framework addresses a central limitation of conventional CoT distillation: intermediate hardware decisions are typically represented as free-form text and receive little direct supervision despite their importance to final RTL correctness. By formulating reasoning as a sequence of intermediate states and propagating terminal testbench feedback through MCTS, MCTS-SFT provides a mechanism for discovering and valuing these otherwise non-verifiable design decisions.

Our controlled experiments show that the strongest performance arises from the combination of structured hardware planning, verifier-guided search, and explicit preservation of the searched plan during SFT. Structured planning is independently valuable, while Plan-MCTS further improves functional trace discovery beyond one-shot planning and unguided tree exploration. Moreover, when the underlying trajectories and Verilog outputs are held fixed, retaining the explicit structured plan produces substantially stronger downstream performance than either omitting the plan or rendering it as less-structured natural language. In the larger 6k-trace evaluation, Plan-MCTS also improves Verilog correctness over full-response CoT distillation while reducing downstream generation length by approximately 50\% on VerilogEval-v2.

These results indicate that effective reasoning supervision depends not only on whether a trajectory reaches a correct final program, but also on which intermediate design states are discovered and how those states are represented during training. This suggests a broader direction for hardware reasoning models in which executable feedback is used to shape intermediate design representations that are both functionally grounded and effective for downstream learning.

\section*{AI Use Statement}

Large language models are a central component of the research methodology in this work. We use LLMs as teacher models to generate, expand, and curate reasoning trajectories and Verilog implementations that form the training datasets evaluated in our experiments. LLM-based tools were also used in a limited capacity during manuscript preparation to assist with formatting and editorial refinement. All scientific claims, experimental design choices, analyses, figures, tables, and conclusions were determined by the authors.

\section*{Ethics Statement}

This work studies methods for improving language-model-based Verilog generation using automatically generated reasoning traces and functional verification. The experiments use existing research datasets, benchmarks, and model-generated training data and do not involve human subjects or private personal information. As with other LLM code-generation systems, generated RTL may contain functional or implementation errors if used without appropriate verification. Our methodology incorporates executable testing during data curation, but such verification should not be interpreted as guaranteeing correctness or suitability for deployment in safety- or security-critical hardware.

\section*{Reproducibility Statement}

We provide the methodological and experimental details needed to reproduce the main results throughout the paper and appendices. Section~\ref{sec:framework} defines the MCTS-SFT and Plan-MCTS formulations, while Appendix~\ref{app:mcts_details} provides additional MCTS algorithm and state-representation details. The experimental setup specifies the controlled and extended-scale evaluation protocols, and Appendix~\ref{app:implementation} reports the available curation, training, model, and evaluation configurations and hyperparameters. Additional ablations and generalization experiments are included in the appendices to clarify the conditions under which the reported findings were obtained. We also document the distinctions between teacher configurations, search strategies, and SFT-target representations so that the controlled comparisons can be reconstructed consistently.

\section*{Acknowledgments}
The authors acknowledge the support from the Purdue Center for Secure Microelectronics Ecosystem - CSME\#210205.

\bibliography{conference}
\bibliographystyle{conference}

\newpage
\appendix
\section{Appendix}
\subsection{Implementation and Experimental Configuration}
\label{app:implementation}

This appendix provides the implementation and hyperparameter details omitted from the main text. Because the controlled 1k-trace study and the extended 6k-trace study use different teacher configurations, we describe them separately.

\subsubsection{Reasoning-Trace Curation}
\label{app:curation_config}

\paragraph{Extended 6k-trace study.}
This experiment uses 600 natural-language specifications sampled from the VeriThoughts training set and generates 10 trajectories per specification, yielding 6,000 traces for each full curation strategy. For MCTS, the maximum number of search iterations per specification is 10 and the maximum number of child nodes generated at an expansion is 3. Intermediate reasoning-node generations are limited to 2,048 tokens, while terminal rollouts are allowed up to 16,184 tokens. Curation-time generation uses temperature 1.0.

The functional reward uses automatically constructed testbenches adapted from the CodeV-R1~\citep{zhu2025qimeng} verification open-source material. CoT and the alternative trace-generation strategies are sampled 10 times per specification to match the number of generated trajectories. The answer-only baseline is obtained by post-processing CoT examples to retain only the final answer, while the Best-of-$N$ Plan-MCTS baseline contains 600 traces selected according to average node score.

\begin{table}[h]
\centering
\small
\caption{Curation configuration reported for the extended 6k-trace study.}
\label{tab:curation_hparams}
\begin{tabular}{l l}
\toprule
\textbf{Parameter} & \textbf{Setting} \\
\midrule
Source specifications & 600 VeriThoughts prompts \\
Trajectories per specification & 10 \\
Full dataset size & 6,000 traces \\
Maximum MCTS iterations & 10 \\
Maximum child nodes per expansion & 3 \\
Intermediate-node token limit & 2,048 \\
Terminal-rollout token limit & 16,184 \\
Curation temperature & 1.0 \\
Best-of-$N$ dataset size & 600 traces \\
Best-of-$N$ selection criterion & Average node score \\
\bottomrule
\end{tabular}
\end{table}

\paragraph{Teacher models.}
In the extended study, DeepSeek-R1 is used for CoT generation, Thought-MCTS, and the reasoning rollouts used by Plan-MCTS. DeepSeek-V3 is used to generate the intermediate non-reasoning Plan-MCTS planning states. All experiments access these models through OpenRouter.

 DeepSeek-V3 was used for intermediate Plan-MCTS updates to reduce expansion cost. An LLM is invoked during each expansion, while only the compact plan decision is retained from intermediate or unselected branches. DeepSeek-R1 is then used to materialize selected plans into reasoning and final Verilog.

\paragraph{Controlled 1k-trace study.}
The controlled study uses 100 specifications sampled from the larger 600-specification set and generates 10 traces per specification, yielding 1,000 examples for each strategy. To remove the teacher-model confound, DeepSeek-R1 is used for every LLM call, including CoT generation, Thought-MCTS expansion and rollout, Plan-CoT planning and rollout, Plan-RandomTree expansion and rollout, and Plan-MCTS expansion and rollout. All controlled methods use the same reward, inference settings, generation budget, student model, and SFT configuration.

\begin{table}[t]
\centering
\small
\caption{Controlled 1k-trace study configuration.}
\label{tab:controlled_config}
\begin{tabular}{l l}
\toprule
\textbf{Parameter} & \textbf{Setting} \\
\midrule
Source specifications & 100 VeriThoughts prompts \\
Trajectories per specification & 10 \\
Dataset size per strategy & 1,000 traces \\
Teacher model & DeepSeek-R1 \\
Teacher assignment & All LLM calls and stages \\
Downstream benchmark & VerilogEval-v1-Human \\
Generations per evaluation problem & $n=10$ \\
\bottomrule
\end{tabular}
\end{table}

\subsubsection{Supervised Fine-Tuning}
\label{app:sft_config}

All experiments use LlamaFactory framework scripts open-sourced via VeriThoughts~\citep{yubeaton2025verithoughts} for supervised fine-tuning. All compared datasets use the same optimization configuration: a learning rate of $5\times10^{-5}$, maximum sequence length of 16,384 tokens, three training epochs, and LoRA with rank 64. Training is performed on a single NVIDIA H200 GPU.

\begin{table}[t]
\centering
\small
\caption{Supervised fine-tuning configuration.}
\label{tab:sft_hparams}
\begin{tabular}{l l}
\toprule
\textbf{Parameter} & \textbf{Setting} \\
\midrule
Training framework & LlamaFactory \\
Learning rate & $5\times10^{-5}$ \\
Maximum sequence length & 16,384 tokens \\
Training epochs & 3 \\
Adaptation method & LoRA \\
LoRA rank & 64 \\
Training hardware & 1 NVIDIA H200 GPU \\
\bottomrule
\end{tabular}
\end{table}

The controlled experiments retain the same SFT configuration for all compared datasets. The student LLM used in this study is Qwen2.5-Coder-7B-Instruct.

\subsubsection{Downstream Evaluation}
\label{app:evaluation_config}

For the extended 6k-trace study, downstream evaluation is performed on VerilogEval-v1-Human and the VerilogEval-v2 Code Completion and Spec-to-RTL tasks. Generation is limited to 8,192 tokens, with temperature 0.5 and top-$p=0.9$. Each benchmark problem is sampled $n=20$ times. The final complete Verilog module is extracted from each model response using the same post-processing procedure across all evaluated models.

\begin{table}[t]
\centering
\small
\caption{Downstream generation configuration reported for the extended study.}
\label{tab:evaluation_hparams}
\begin{tabular}{l l}
\toprule
\textbf{Parameter} & \textbf{Setting} \\
\midrule
Benchmarks & VerilogEval-v1-Human, VerilogEval-v2 \\
VerilogEval-v2 tasks & Code Completion, Spec-to-RTL \\
Maximum generation length & 8,192 tokens \\
Temperature & 0.5 \\
Top-$p$ & 0.9 \\
Generations per problem & $n=20$ \\
Output processing & Consistent final-module extraction \\
\bottomrule
\end{tabular}
\end{table}

For the controlled 1k-trace study, downstream comparisons are performed on VerilogEval-v1-Human with $n=10$ generations per problem. 
\subsection{Additional Algorithm and Formulation Details}
\label{app:mcts_details}

This appendix provides additional details of the verifier-guided search procedure described in Section~\ref{sec:framework}. We retain the notation introduced in the main text: for a specification $x$, an intermediate state is $S_t=(x,h_t)$, where $h_t$ denotes either a free-form reasoning state or a structured hardware plan. The state transition, rollout, verifier reward, UCT selection rule, and action-value update are defined in Section~\ref{sec:verifier_search}; here we provide the remaining search mechanics and the full formulation of the Thought-MCTS comparison.

\subsubsection{MCTS Search Procedure}
\label{app:mcts_algorithm}

For each specification, search begins from the root
\begin{equation}
    S_0=(x,h_0),
\end{equation}
where $h_0$ is an empty intermediate representation. The search terminates when the current state is judged ready for Verilog generation or when a maximum search depth $D$ is reached. For each specification, MCTS performs $M$ simulations, each consisting of selection, expansion, rollout, and backpropagation.

\paragraph{Selection.}
Starting from $S_0$, search recursively selects among previously expanded actions using the UCT rule defined in Eq.~\ref{eq:uct}. For a state $S$, $\mathcal{A}_{\mathrm{tree}}(S)$ denotes the set of actions that have already been expanded, $N(S)$ is the number of visits to the state, and $N(S,a)$ is the number of times action $a$ has been selected from $S$.

We use UCT rather than a prior-weighted PUCT rule because the actions in our setting are variable-length reasoning or planning generations. Calibrated action priors over these multi-token outputs are not available from the teacher models and are not directly comparable across candidate actions.

\paragraph{Expansion.}
When selection reaches an expandable leaf state
$S_{\ell}=(x,h_{\ell})$, the expansion model proposes $K$ candidate actions:
\begin{equation}
    \left\{
        a_{\ell}^{(1)},\ldots,a_{\ell}^{(K)}
    \right\}
    \sim
    \pi_{\theta}^{\mathrm{expand}}
    (\cdot \mid S_{\ell}).
    \label{eq:appendix_expansion}
\end{equation}
Each candidate defines a successor state
\begin{equation}
    S_{\ell+1}^{(k)}
    =
    T(S_{\ell},a_{\ell}^{(k)})
    =
    \left(x,
    h_{\ell}\oplus a_{\ell}^{(k)}
    \right),
    \label{eq:appendix_child}
\end{equation}
where $\oplus$ is representation-specific. For Thought-MCTS, the candidate action is a free-form reasoning step; for Plan-MCTS, it is a structured update to the hardware plan. One newly created child is then selected for rollout.

\paragraph{Rollout and verification.}
The selected child is passed to the rollout model, which completes the reasoning process and generates final Verilog as described in Eq.~\ref{eq:rollout}. The resulting implementation is evaluated using the terminal verifier reward in Eq.~\ref{eq:reward}. The terminal state therefore associates the specification and intermediate search representation with a materialized reasoning trace and final Verilog implementation.

\paragraph{Backpropagation.}
The terminal verifier reward is propagated along every state--action edge visited during the simulation. In addition to the running action-value update in Eq.~\ref{eq:q_update}, the visitation counts are updated as
\begin{align}
    N(S) &\leftarrow N(S)+1, \\
    N(S,a) &\leftarrow N(S,a)+1.
\end{align}
Consequently, functional feedback observed only after producing complete Verilog contributes to the value estimates of the intermediate reasoning or planning actions that led to that implementation.

\subsubsection{Thought-MCTS: Free-Form Reasoning Search}
\label{app:thought_mcts}

Thought-MCTS applies the same selection, rollout evaluation, and backpropagation procedure as Plan-MCTS, but directly represents the search state as accumulated natural-language reasoning. After $t$ reasoning decisions,
\begin{equation}
    h_t^{\mathrm{thought}}
    =
    (z_1,z_2,\ldots,z_t),
    \label{eq:thought_state}
\end{equation}
where each $z_i$ is an LLM-generated reasoning segment. An expansion action proposes the next reasoning segment,
\begin{equation}
    a_t^{\mathrm{thought}}
    =
    z_{t+1}
    \sim
    \pi_{\theta}^{\mathrm{expand}}
    (\cdot \mid x,z_1,\ldots,z_t),
    \label{eq:thought_action}
\end{equation}
and the transition appends it to the existing prefix:
\begin{equation}
    h_{t+1}^{\mathrm{thought}}
    =
    h_t^{\mathrm{thought}} \circ z_{t+1},
    \label{eq:thought_transition}
\end{equation}
where $\circ$ denotes sequence concatenation.

During expansion, candidate reasoning steps are generated conditioned on the specification and accumulated prefix. During rollout, the same prefix conditions generation of the remaining reasoning and final Verilog:
\begin{equation}
    (r_T,v_T)
    \sim
    \pi_{\theta}^{\mathrm{rollout}}
    (\cdot \mid x,z_1,\ldots,z_t).
    \label{eq:thought_rollout}
\end{equation}
Thus, Thought-MCTS changes only the representation searched by MCTS: intermediate states consist directly of natural-language reasoning rather than explicit hardware-design fields. Because this text is retained in later expansion and rollout contexts, longer reasoning prefixes also increase the amount of intermediate context carried through search.

\subsubsection{Additional Plan-MCTS Search Details}
\label{app:plan_mcts_details}

As defined in Section~\ref{sec:plan_mcts}, Plan-MCTS instead represents $h_t$ as a partial hardware plan $p_t$ and each action as a structured update $\Delta p_t$. The expansion model may determine which plan field to update based on the current state and may append additional decisions to fields that require further specification. As a result, multiple search steps may refine the same aspect of the design rather than requiring each field to be populated exactly once.

Candidate plan updates are otherwise processed using the same selection, rollout, verification, and backpropagation procedure described above. During rollout, the accumulated plan is supplied as design context to the rollout model, which materializes it into natural-language reasoning and final Verilog:
\begin{equation}
    (r_T,v_T)
    \sim
    \pi_{\theta}^{\mathrm{rollout}}
    (\cdot \mid x,p_T).
\end{equation}
The verifier evaluates only $v_T$, and the resulting terminal reward is propagated through the plan-update actions encountered along the selected search trajectory.

\subsection{Additional Controlled Analyses}
\label{app:additional_controlled}

\subsubsection{Compression versus Hardware-Specific Structure}
\label{app:summary_mcts}

The controlled experiments in Section~\ref{sec:results_rq2} show that preserving the structured hardware plan substantially improves downstream SFT. To determine whether this effect can instead be explained by reducing the length of the intermediate search state, we additionally evaluate \textbf{Summary-MCTS}.

Summary-MCTS is identical to Thought-MCTS except that, after each generated thought, an additional LLM call condenses the accumulated reasoning before the next expansion. It therefore provides a compression control without introducing the hardware-specific interface, timing, state, behavior, or implementation fields used by Plan-MCTS.

\begin{table}[h]
\centering
\small
\caption{
Comparison of reasoning-state compression and hardware-specific structure in the controlled study.
}
\label{tab:summary_mcts}
\begin{tabular}{l r r r}
\toprule
\textbf{Method / representation}
& \textbf{pass@1}
& \textbf{pass@5}
& \textbf{pass@10} \\
\midrule
Thought-MCTS, full free-form prefix
& 27.37
& 46.71
& 51.28 \\

Summary-MCTS, compressed free-form prefix
& 29.23
& 46.43
& 51.92 \\

Plan-MCTS, rollout only
& 29.17
& 46.48
& 52.56 \\

Plan-MCTS + rendered NL plan
& 30.0
& 47.6
& 55.1 \\

Plan-MCTS + explicit structured plan
& 31.5
& 50.8
& 57.7 \\
\bottomrule
\end{tabular}
\end{table}

Summarization modestly improves Thought-MCTS at pass@1 by 1.86 points and at pass@10 by 0.64 points, but does not reproduce the gains obtained from explicit Plan-MCTS supervision. The explicit structured-plan target exceeds Summary-MCTS by 2.27/4.37/5.78 points at pass@1/5/10. These results indicate that compression alone is insufficient to explain the benefit of the structured hardware representation.

The natural-language and explicit-plan targets provide a complementary comparison. Both are derived from the same Plan-MCTS trajectories and underlying Verilog implementations, but retaining the explicit structured representation improves pass@1/5/10 by 1.5/3.2/2.6 points over the rendered natural-language plan. Thus, the organization of the intermediate hardware decisions also affects their effectiveness as SFT supervision.

\subsection{Additional Generalization Experiments}
\label{app:generalization}

Beyond the controlled and extended-scale evaluations, we conducted two additional studies to examine whether the observed benefits extend across teacher models and larger training sets.

\subsubsection{Generalization to a Second Teacher Model}
\label{app:second_teacher}

To examine dependence on the teacher model, we additionally use GPT-5.4-mini for reasoning-trace curation while retaining the same Qwen student. Table~\ref{tab:second_teacher} reports downstream performance for Thought-MCTS and Plan-MCTS.

\begin{table}[h]
\centering
\small
\caption{
Generalization to GPT-5.4-mini as the curation teacher.
}
\label{tab:second_teacher}
\begin{tabular}{l r r r}
\toprule
\textbf{Teacher strategy}
& \textbf{pass@1}
& \textbf{pass@5}
& \textbf{pass@10} \\
\midrule
Thought-MCTS
& 33.21
& 45.90
& 50.64 \\

Plan-MCTS
& 33.78
& 46.21
& 50.00 \\
\bottomrule
\end{tabular}
\end{table}

Both search representations transfer to the second teacher. Plan-MCTS is slightly stronger at pass@1 and pass@5, while Thought-MCTS is slightly stronger at pass@10. The small differences do not establish a consistent ranking between the two variants, but indicate that the verifier-guided curation framework is not restricted to the teacher configuration.

\begin{table}[h]
\centering
\small
\caption{
Scaling standard Thought-MCTS supervision to larger training datasets.
}
\label{tab:mcts_scaling}
\begin{tabular}{l r r r}
\toprule
\textbf{Training configuration}
& \textbf{pass@1}
& \textbf{pass@5}
& \textbf{pass@10} \\
\midrule
Qwen2.5 baseline
& 25.6
& 38.9
& 42.6 \\

SFT, 10k MCTS traces
& 38.2
& 51.8
& 56.6 \\

SFT, 15k MCTS traces
& 41.7
& 54.9
& 59.6 \\

SFT, 25k MCTS traces
& 43.7
& 54.1
& 58.0 \\

SFT, 40k MCTS traces
& 45.6
& 56.0
& 58.9 \\

SFT, 60k MCTS traces
& 49.0
& 60.0
& 63.3 \\
\bottomrule
\end{tabular}
\end{table}

\subsubsection{Scaling MCTS-Curated Supervision}
\label{app:data_scaling}

We additionally examine whether MCTS-curated supervision continues to benefit the student as the training dataset grows. For this experiment, we scale the standard Thought-MCTS pipeline to 6,000 unique prompts and up to 60,000 reasoning traces. Table~\ref{tab:mcts_scaling} reports the resulting pass@$k$ performance.

Pass@1 increases consistently with dataset scale, reaching 49.0 at 60k traces, corresponding to a 23.4-point improvement over the base model and a 10.8-point improvement over the 10k-trace model. Pass@5 and pass@10 also improve substantially overall, although the intermediate results are not strictly monotonic. These results provide evidence that the benefits of MCTS-curated supervision persist at substantially larger prompt and dataset scales.

\end{document}